\documentclass[fleqn]{revtex4}

\usepackage{amsmath,graphicx}
\usepackage{verbatim,dcolumn}
\usepackage{multirow}
\usepackage[symbol]{footmisc}

\def\br{{\mathbf{r}}}
\def\bp{{\mathbf{p}}}
\def\bq{{\mathbf{q}}}
\def\bL{{\mathbf{L}}}
\def\bI{{\mathbf{I}}}
\def\bs{{\mathbf{s}}}
\def\bE{{\mathbf{E}}}
\def\bB{{\mathbf{B}}}
\def\bA{{\mathbf{A}}}
\def\F{{\overline{F}}}
\def\bN{{\boldsymbol{\nabla}}}
\def\bpi{{\boldsymbol{\pi}}}
\def\bsigma{{\boldsymbol{\sigma}}}

\begin{document}

\title{Higher-order corrections to spin-orbit and spin-spin tensor interactions in hydrogen molecular ions: theory and application to H$_2^+$}

\author{Mohammad Haidar$^1$\footnote[1]{Present affiliations: 1. Sorbonne Universit\'e, Laboratoire de Chimie Th\'eorique (LCT), 4 place Jussieu, F-75005 Paris, France 2. Sorbonne Universit\'e, CNRS, Universit\'e Paris Cit\'e, Laboratoire Jacques-Louis Lions (LJLL), 4 place Jussieu, F-75005 Paris, France 3. TotalEnergies, Tour Coupole, 2 Pl. Jean Millier, F-92078 Paris la D\'efense, France}, Vladimir I. Korobov$^2$, Laurent Hilico$^{1,3}$, and Jean-Philippe Karr$^{1,3}$}
\affiliation{$^1$Laboratoire Kastler Brossel, Sorbonne Universit\'e, CNRS, ENS-Universit\'e PSL, Coll\`ege de France, 4 place Jussieu, F-75005 Paris, France}
\affiliation{$^2$Bogoliubov Laboratory of Theoretical Physics, Joint Institute for Nuclear Research, Dubna 141980, Russia}
\affiliation{$^3$Universit\'e d'Evry-Val d'Essonne, Universit\'e Paris-Saclay, Boulevard Fran\c cois Mitterrand, F-91000 Evry, France}

\begin{abstract}
We consider higher-order corrections to hyperfine coefficients related to the spin-orbit and spin-spin tensor interactions in hydrogen molecular ions. The $m\alpha^7 \ln(\alpha)$-order radiative correction is derived in the NRQED framework. We present complete numerical calculations, including as well the $m\alpha^6$-order relativistic correction, for the case of H$_2^+$. The theoretical uncertainty is reduced by more than one order of magnitude with respect to the Breit-Pauli level, down to a few ppm. We also compare our results with available rf spectroscopy data.
\end{abstract}

\maketitle

\section{Introduction}

In recent years, precision spectroscopy of hydrogen molecular ions has established itself as a fruitful direction for fundamental metrology. Rovibrational transition frequencies in HD$^+$ have been measured with very high accuracies~\cite{Alighanbari20,Patra20} and compared with theoretical predictions~\cite{Korobov21} to obtain improved determinations of the proton-electron mass ratio or constrain hypothetical interactions beyond the Standard Model~\cite{Germann21}. In these works, accurate predictions of the hyperfine structure have been used to extract a ``spin-averaged'' transition frequency from the measured hyperfine components. Discrepancies between theory and experiments have been observed in the hyperfine slitting of the rovibrational lines~\cite{Korobov20,Karr20}, which in some cases increases the uncertainty of rovibrational transition frequencies~\cite{Patra20,Koelemeij22}. This makes it highly desirable to improve further the hyperfine structure theory in hydrogen molecular ions.

The theory of the leading hyperfine interaction, namely the ``Fermi'' spin-spin contact interaction that gives rise to the main ($\sim 1$~GHz) splitting in HD$^+$ and \textit{ortho}-H$_2^+$, has been recently improved~\cite{Karr20,Korobov16}. The next step consists in improving the next largest hyperfine coefficients, related to the electronic spin-orbit and spin-spin tensor (dipolar) interactions~\cite{Korobov06,Bakalov06}. It is worth noting that the spin-orbit and spin-spin tensor interactions, being essentially free of nuclear finite-size and structure corrections, allow for more precise tests of the theory with respect to the contact interaction. In Ref.~\cite{Korobov20}, we derived the effective Hamiltonian for relativistic corrections of order $m\alpha^6$ in the hydrogen molecular ions, following the nonrelativistic QED (NRQED) approach that had been previously validated by applying it to the hyperfine splitting of the 2P state in hydrogen~\cite{Haidar20}. This allowed us to get improved values of the spin-orbit coefficient for a few states~\cite{Korobov20}. In this work, we improve the theory further by deriving the radiative correction at the following order $m\alpha^7 \ln{\alpha}$.

We then present extensive numerical calculations of the spin-orbit and spin-spin tensor coefficients in the slightly simpler case of H$_2^+$, whereas HD$^+$ will be considered in a forthcoming publication. There are several motivations to study the hyperfine structure specifically in H$_2^+$. Recent efforts and proposals towards high-resolution laser spectroscopy of this ion~\cite{Chou17,Schmidt20,Tu21} offer new opportunities to test the theory; accurate theoretical predictions of the hyperfine splitting are also likely to be required to extract spin-averaged transition frequencies, similarly to HD$^+$~\cite{Alighanbari20,Patra20}. Moreover, H$_2^+$ is of high astrophysical importance due to its role in the formation of H$_3^+$. This has made its radio-astronomical detection, using e.g. hyperfine transitions analogous to the 21-cm line in atomic hydrogen, a long-standing goal~\cite{Jefferts70,Shuter86,Black98}. Interest in H$_2^+$ is further enhanced by prospects of experimental studies on the antihydrogen molecular ion $\bar{H}_2^-$, aimed at performing improved tests of the CPT symmetry~\cite{Myers18}; some of these tests could be performed through measurements of hyperfine-Zeeman transitions. Finally, a few hyperfine intervals that are essentially independent from the ``Fermi'' coefficients have been measured with very high precision ($\sim 10^{-7}$)~\cite{Menasian73}, thus providing a stringent test of theory for the spin-orbit and spin-spin tensor interactions.

\section{NRQED Hamiltonian}

For calculation of $m\alpha^7$-order corrections, a more complete version of the NRQED Hamiltonian used in our previous works~\cite{Korobov20,Haidar20} is required. Namely, its coefficients should be determined up to first order in $\alpha$ by matching NRQED and QED scattering amplitudes~\cite{Kinoshita96,Manohar97,Hill13}. Writing only the terms that are relevant for the present consideration, the NRQED Hamiltonian has the form
\begin{equation}\label{HamiltonianGI}
\begin{array}{@{}l}\displaystyle
H_{I} = eA_0 + \frac{\bpi^2}{2m} - \frac{\bpi^4}{8m^3} - c_F\frac{e}{2m}\bsigma\!\cdot\!\bB
   -c_D\frac{e}{8m^2}[\boldsymbol{\partial}\!\cdot\!\bE]
   +c_S\frac{e}{8m^2}\,\bsigma\!\cdot\!
      \Bigl(
         \bpi\!\times\!\bE-\bE\!\times\!\bpi
      \Bigr)
\\[3mm]\hspace{8mm}\displaystyle
   +\,c_{W}\frac{e}{8m^3}\Bigl\{\bpi^2,\bsigma\!\cdot\!\bB\Bigr\}
   - c_{q^2}\frac{e}{8m^3} \bsigma\!\cdot\![\Delta\bB]
   +c_{p'p}\frac{e}{8m^3}
         \Bigl( \bsigma\!\cdot\!\bpi\>\bB\!\cdot\!\bpi + \bpi\!\cdot\!\bB\>\bsigma\!\cdot\!\bpi \Bigr)
   +c_M\frac{e}{8m^3}
         \Bigl\{
            \bpi,[\boldsymbol{\partial}\!\times\!\bB]
         \Bigr\}
\\[3mm]\hspace{8mm}\displaystyle
   +\,c_{X_1}\frac{ie}{128m^4}\left[\bpi^2,(\bpi\!\cdot\!\bE+\bE\!\cdot\!\bpi)\right]
   +c_{X_2}\frac{e}{64m^4}\,\Bigl\{\bpi^2,\left[\boldsymbol{\partial}\!\cdot\!\bE\right]\Bigr\}
   -c_{X_3}\frac{e}{8m^4}\,\Bigl[\Delta\left[\boldsymbol{\partial}\!\cdot\!\bE\right]\Bigr]
\\[3mm]\hspace{8mm}\displaystyle
   -\,c_{Y_1}\frac{e}{64m^4}
      \Bigl\{
        \bpi^2,\bsigma\!\cdot\!
         \Bigl(\bpi\!\times\!\bE\!-\!\bE\!\times\!\bpi\Bigr)
      \Bigr\}
   +c_{Y_2}\frac{ie}{4m^4}\>\epsilon_{ijk}\sigma^i \pi^j[\boldsymbol{\partial}\!\cdot\!\bE]\pi^k\,
\end{array}
\end{equation}
where $\bpi=\bp-e\bA$, $\bE=-\partial_t\bA\!-\!\bN A_0$, $\bB = \boldsymbol{\nabla}\!\times\!\bA$, $[X,Y] = XY - YX$, and $\{X,Y\} = XY + YX$. Square brackets around quantities imply that derivatives act only within the bracket (this notation only applies to Eq.~(\ref{HamiltonianGI}) and is no longer used in the following).

The above expression differs from Eq.~(1) of~\cite{Hill13}, which is complete up to the order $1/m^4$, in several details. Firstly, we have omitted terms involving the coefficients $c_{X_4}$ and $c_{X_7}-c_{X_{12}}$, which only contribute at orders $m\alpha^8$ and above, and the two-photon (seagull) terms involving $c_{A_1}$ and $c_{A_2}$, since for the corrections we aim to calculate it is sufficient to perform a matching of one-photon scattering amplitudes. Secondly, the terms involving the coefficients $c_{W_1}$ and $c_{W_2}$ have been reformulated by introducing the coefficients $c_W$ and $c_q^2$, as done in~\cite{Kinoshita96,Nio97}. In a similar way, we have reformulated the terms involving $c_{X_5}$ and $c_{X_6}$ by introducing $c_{Y_1}$ and $c_{Y_2}$. The reason behind these transformations is to get simpler expressions for the NRQED effective potentials. Finally, for convenience we have changed the definitions of $c_{X_1} - c_{X_3}$ by introducing numerical prefactors in the corresponding terms.

For the following calculations, $\bpi$ can be replaced by $\bp$ in the last three lines because the terms involving $\bA$ only contribute at higher orders.

The QED scattering amplitude at tree level for a static scalar field $A_0(\bq)$ is
\begin{equation} \label{scalarscat}
A_E^{\rm QED} (p,p') = -i A_0 J^0 (p,p') \,,
\end{equation}
where $p$, $p'$ are the four-momenta of the incident and scattered particle, and $J$ is the charge-current density operator, which is written in terms of the Dirac and Pauli form factors $F_1 (q^2)$ and $F_2 (q^2)$ (with $\bq = \bp' - \bp$):
\begin{equation} \label{current}
J^{\mu} = ie \, \overline{u}(p') \left( \gamma^{\mu} F_1 (q^2) + \frac{i\kappa}{2m} \sigma^{\mu\nu} q_{\nu} F_2 (q^2) \right) u(p).
\end{equation}
Here, $\kappa$ is the particle's anomalous magnetic moment, and $u(p)$, $u(p')$ are on-shell Dirac spinors. Using the nonrelativistic normalization condition $u^{\ast}(p) u(p) = 1$, a Dirac spinor can be expressed in terms of a Schr\"odinger-Pauli spinor $\psi(p)$ as
\begin{equation}
u(p) = \sqrt{\frac{E_p + m}{2 E_p}} \begin{pmatrix} \psi(p) \\
          \frac{\bsigma\cdot\bp}{E_p + m} \psi(p)
          \end{pmatrix} \,,
          E_p = \sqrt{m^2 + \bp^2} \,.
\end{equation}
It can then be expanded in powers of $\bp^2/m^2$:
\begin{equation}
u(p) \approx \begin{pmatrix} \label{spinorexp}
\left[ 1 - \frac{\bp^2}{8m^2} + \frac{11\bp^4}{128m^4} + \ldots \right] \psi \\
\frac{\bsigma\cdot\bp}{2m} \left[  1 - \frac{3\bp^2}{8m^2} + \ldots \right] \psi
\end{pmatrix}
\end{equation}
The form factors may also be expanded as
\begin{eqnarray}
F_1 (q^2) &=& \F_1 - \F'_1 \frac{\bq^2}{m^2} + \F''_1 \frac{\bq^4}{m^4} + \ldots \\
F_2 (q^2) &=& \F_2 - \F'_2 \frac{\bq^2}{m^2} + \ldots \label{ffexp}
\end{eqnarray}
with $\F_1 = \F_2 = 1$ for an electron. Using Eqs.~(\ref{scalarscat})-(\ref{current}) and~(\ref{spinorexp})-(\ref{ffexp}), one gets the following expansion of the QED scattering amplitude:
\begin{eqnarray}
A_E^{\rm QED} (p,p') &=& \psi^{\ast}(p') e A_0 \left[ \F_1 -\frac{\bq^2}{8m^2} (\F_1 + 2\kappa \F_2 + 8 \F'_1) + i \frac{\bsigma\!\cdot\! \left(\bq\!\times\!\bp\right)}{4m^2} (\F_1 + 2\kappa \F_2) \right. \nonumber \\
&& \left. + \frac{\bq^4}{8m^4} (\F'_1 + 2\kappa \F'_2 + 8 \F''_1)  - i \frac{\bsigma\!\cdot\! \left(\bq\!\times\!\bp\right) \bq^2}{4m^4} (\F'_1 + 2\kappa \F'_2) + \frac{\bq^2 (p'^2+p^2)}{64m^4} (3 \F_1 + 4 \kappa \F_2) \right. \nonumber \\
&& \left. + \frac{(p'^2-p^2)^2}{128m^4} (5 \F_1 + 4\kappa \F_2) -i \frac{\bsigma\!\cdot\! \left( \bq\!\times\!\bp \right) \left( p'^2 + p^2 \right)}{32m^4} (3 \F_1 + 4 \kappa \F_2) + \ldots \right] \psi (p). \label{scalarQED}
\end{eqnarray}
Similarly, for a vector field $\bA(\bq)$, the scattering amplitude
\begin{equation}
A_M^{\rm QED} (p,p') = -i A_{\mu} J^{\mu} (p,p')
\end{equation}
can be expanded as follows:
\begin{eqnarray}
A_M^{\rm QED} (p,p') &=& \psi^{\ast}(p') e \bA \!\cdot\! \left[ -\frac{\left( \bp'+\bp \right)}{2m} \F_1 -i \frac{\left( \bsigma\!\times\!\bq \right)}{2m} (\F_1 + \kappa \F_2) + \frac{\bq^2 (\bp' + \bp)}{16m^3} (8 \F'_1 + \kappa \F_2) \right. \nonumber \\
&& + \frac{(\bp' + \bp)(p'^2+p^2)}{8m^3} \F_1 + \frac{\bq \left( p'^2 - p^2 \right)}{16m^3} (\F_1 - \kappa \F_2) \label{vectorQED} \\
&&  +i \frac{\left( \bsigma\!\times\!\bq \right) \left( p'^2 + p^2 \right)}{8m^3} \F_1 +i \frac{\left[ \bsigma\!\times\!(\bp' + \bp) \right] \left( p'^2 - p^2 \right)}{16m^3} \F_1 \nonumber \\
&& \left. + i \frac{(\bsigma\!\cdot\!\bp') \left( \bp\!\times\!\bq \right) + (\bsigma\!\cdot\!\bp) \left( \bp'\!\times\!\bq \right)}{8m^3} \kappa \F_2 + i \frac{\bq^2 \left( \bsigma\!\times\!\bq \right)}{16m^3} (\kappa \F_2 + 8 \F'_1 + 8 \kappa \F'_2) \right] \psi (p). \nonumber
\end{eqnarray}
The NRQED scattering amplitude is directly obtained from the Hamiltonian~(\ref{HamiltonianGI}). For a scalar field one gets
\begin{eqnarray}
A_E^{\rm NRQED} (p,p') &=& \psi^{\ast}(p') e A_0 \left[ 1 - c_D \frac{\bq^2}{8m^2} + i c_S \frac{\bsigma\!\cdot\!\left[\bq\!\times\!\left(\bp'+\bp\right)\right]}{8m^2}  \right. \nonumber \\
&& + c_{X_1} \frac{(p'^2 - p^2)^2}{128m^4} + c_{X_2} \frac{(p'^2 + p^2)\bq^2}{64m^4}   + c_{X_3} \frac{\bq^4}{8m^4}  \nonumber \\
&& \left. - ic_{Y_1} \frac{(p'^2 + p^2) \bsigma\!\cdot\!\left[\bq\!\times\!(\bp'+\bp)\right]}{64m^4}  + i c_{Y_2} \frac{\bsigma\!\cdot\!\left[ \bq\!\times\! (\bp'+\bp) \right] \bq^2}{8m^4}  \right] \psi(p) \,, \label{scalarNRQED}
\end{eqnarray}
and for a vector field:
\begin{eqnarray}
A_M^{\rm NRQED}(p,p') &=& \psi^{\ast}(p') e \bA \!\cdot\! \left[ - \frac{\left( \bp' + \bp \right)}{2m} - i c_F \frac{\left( \bsigma\!\times\!\bq \right)}{2m}  + \frac{(p'^2 + p^2) \left( \bp' + \bp \right)}{8m^3}  \right. \nonumber \\
&& + i c_{W} \frac{\left( p'^2 + p^2 \right) \left( \bsigma\!\times\!\bq \right)}{8m^3}  + i c_{q^2} \frac{\bq^2 \left( \bsigma\!\times\!\bq \right)}{8m^3}  \nonumber \\
&& \left. + i c_{p'p} \frac{\left( \bsigma\!\cdot\!\bp' \right) \left( \bp\!\times\!\bq \right) + \left( \bsigma\!\cdot\!\bp\right) \left( \bp'\!\times\!\bq \right)}{8m^3}
+ c_M \frac{\bq^2 \left( \bp'+\bp \right)}{8m^3}  \right] \psi(p) \,. \label{vectorNRQED}
\end{eqnarray}
Matching Eq.~(\ref{scalarNRQED}) with Eq.~(\ref{scalarQED}) and  Eq.~(\ref{vectorNRQED}) with Eq.~(\ref{vectorQED}) allows us to determine the coefficients of the NRQED Hamiltonian. Note that the last term in the third line of Eq.~(\ref{vectorQED}) does not appear in the corresponding NRQED expression~(\ref{vectorNRQED}), because it is gauge dependent and thus does not contribute to the scattering amplitude. Our final result is:
\begin{align}
c_F &= \F_1 + \kappa \F_2 &
c_D &= \F_1 + 2 \kappa \F_2 + 8 \F'_1 &
c_S &= \F_1 + 2 \kappa \F_2 \nonumber \\
c_{W} &= \F_1 &
c_{q^2} &= \frac{1}{2} (\kappa \F_2 + 8 \F'_1 + 8 \kappa \F'_2) &
c_{p'p} &= \kappa \F_2 \\
c_M &= \frac{1}{2} (\kappa \F_2 + 8 \F'_1) &
c_{X_1} &= 5 \F_1 + 4 \kappa \F_2 &
c_{X_2} &= 3 \F_1 + 4 \kappa \F_2 \nonumber \\
c_{X_3} &= \F'_1 + 2 \kappa \F'_2 + 8 \F''_1 &
c_{Y_1} &= 3 \F_1 + 4 \kappa \F_2 &
c_{Y_2} &= - (\F'_1 + 2 \kappa \F'_2) \nonumber
\end{align}
This can be compared with Ref.~\cite{Hill13} with the help of the relationships
\begin{align}
c_{W} + c_{q^2} &= c_{W_1} & c_{q^2} = c_{W_2} \,, \nonumber \\
c_{Y_1} &= 32 c_{X_5} & - c_{Y_1} + 8 c_{Y_2} &= 32 c_{X_6} \,,
\end{align}
which are easily obtained using the equation $\bq^2 = p'^2 + p^2 - 2\bp'\!\cdot\!\bp$. Our results are in agreement with those of Ref.~\cite{Hill13}, except for $c_{X_1}$ and $c_{X_3}$. Note that these two coefficients do not depend on spin and therefore do not play any role in the interactions studied in this work. For the electron case, the first expansion coefficients of the form factors are:
\begin{equation}
\F'_1 = \frac{\alpha}{3\pi} \left( \ln \left( \frac{m}{\lambda} \right) - \frac{3}{8}  \right) + \ldots, \hspace{15mm}
\F''_1 = \frac{\alpha}{20\pi} \left( \ln \left( \frac{m}{\lambda} \right) - \frac{11}{12} \right) + \ldots, \hspace{15mm}
a_e \F'_2 = \frac{\alpha}{12\pi} + \ldots,
\end{equation}
where $\lambda$ is a photon mass. The coefficients of the NRQED Hamiltonian are then:
\begin{align} \label{NRQED-coeff}
c_F &= 1 + a_e &
c_D &= 1 + 2 a_e + \frac{8}{3} \frac{\alpha}{\pi} \left( \ln \left( \frac{m}{\lambda} \right) - \frac{3}{8} \right)  &
c_S &= 1 + 2 a_e \nonumber \\
c_W &= 1 &
c_{q^2} &= \frac{a_e}{2} + \frac{4}{3} \frac{\alpha}{\pi} \left(  \ln \left( \frac{m}{\lambda} \right) - \frac{1}{8} \right) &
c_{p'p} &= a_e \\
c_M &= \frac{a_e}{2} + \frac{4}{3} \frac{\alpha}{\pi} \left( \ln \left( \frac{m}{\lambda} \right) - \frac{3}{8} \right)&
c_{X_1} &= 5 + 4 a_e &
c_{X_2} &= 3 + 4 a_e \nonumber \\
c_{X_3} &= \frac{\alpha}{\pi} \left( \frac{11}{15} \ln \left( \frac{m}{\lambda} \right) - \frac{13}{40} \right) &
c_{Y_1} &= 3 + 4 a_e &
c_{Y_2} &= -\frac{1}{3} \frac{\alpha}{\pi} \left( \ln \left( \frac{m}{\lambda} \right) + \frac{1}{8} \right) \nonumber
\end{align}
It is important to note that logarithmic contributions can be immediately obtained by substituting the photon mass $\lambda$ in the $\ln(m/\lambda)$ terms with the natural energy scale $m\alpha^2$ (see, e.g.,~\cite{Labelle99}).

\section{Hyperfine structure corrections at orders $m\alpha^6$ and $m\alpha^7\ln(\alpha)$}

\subsection{Terms contributing at the order $m\alpha^6$}

Effective potentials contributing to the spin-orbit and spin-spin tensor interactions can be obtained from the NRQED Hamiltonian, Eq.~(\ref{HamiltonianGI}), using perturbation theory. For the $m\alpha^6$ order, this has been done in our previous work~\cite{Korobov20}. We recall these results before moving on to the new corrections appearing at order $m\alpha^7\ln(\alpha)$. We use natural relativistic units ($\hbar= c = 1$) and the following notations: $\mathbf{s}_e$ is the electron spin, $Z_1,Z_2$ and $M_1,M_2$ are the nuclear charges and masses (here $Z_1 = Z_2 = 1$, $M_1 = M_2 = m_p$), $\br_a = \br_e - \mathbf{R}_a$ ($a=1,2$) is the position of the electron with respect to nucleus $a$, and $\mathbf{p}_e,\mathbf{P}_1,\mathbf{P}_2$ are the impulse operators for the electron and both nuclei, respectively.

We first list the corrections to the electronic spin-orbit interaction. The total energy correction is a sum of first-order and second-order contributions,
\begin{equation}
\Delta E_{so(6)} = \langle H_{so(6)} \rangle + \Delta E_{so(6)}^{2^{nd}-order},
\end{equation}
where $\langle \rangle$ denotes an expectation value with the nonrelativistic wave function $\psi_0$. The $m\alpha^6$-order effective Hamiltonian is
\begin{equation} \label{so6-1st-order}
\begin{array}{@{}l}\displaystyle
H_{so(6)} = c_W \mathcal{U}_W + c_{Y_1} \mathcal{U}_{Y_1} + c_S \mathcal{U}_{CM} + \mathcal{U}_{MM_N},
\\[3mm]\displaystyle
\mathcal{U}_{W} =
   \frac{Z_a}{4m^3M_a}
   \left\{p_e^2,\frac{1}{r_a^3}\bigl(\mathbf{r}_a\!\times\!\mathbf{P}_a\bigr)\right\}\!\cdot\!\mathbf{s}_e,
\\[3mm]\displaystyle
\mathcal{U}_{Y_1} =
   -\frac{Z_a}{16m^4}
         \left\{p_e^2,\frac{1}{r_a^3}(\mathbf{r}_a\!\times\!\mathbf{p}_e)\right\}\!\cdot\!\mathbf{s}_e,
\\[3mm]\displaystyle
\mathcal{U}_{CM} =
   \frac{Z_a^2}{4m^2M_a}\>
     \frac{1}{r_a^4}\left(\mathbf{r}_a\!\times\!\mathbf{P}_a\right)\!\cdot\!\mathbf{s}_e
   +\frac{Z_1Z_2}{4m^2M_1}\>
     \frac{1}{r_1r_2^3}\left(\mathbf{r}_2\!\times\!\mathbf{P}_1\right)\!\cdot\!\mathbf{s}_e
   +\frac{Z_1Z_2}{4m^2M_2}\>
     \frac{1}{r_1^3r_2}\left(\mathbf{r}_1\!\times\!\mathbf{P}_2\right)\!\cdot\!\mathbf{s}_e
\\[3mm]\displaystyle\hspace{10mm}
   -\frac{Z_1Z_2}{4m^2M_a}\>
     \frac{1}{r_1^3r_2^3}\left(\mathbf{r}_1\!\times\!\mathbf{r}_2\right)(\mathbf{r}_a\!\cdot\!\mathbf{P}_a)\!\cdot\!\mathbf{s}_e,
\\[3mm]\displaystyle
\mathcal{U}_{MM_N} =
   -\frac{Z_a^2}{2m^2M_a}\>
      \frac{1}{r_a^4}\left(\mathbf{r}_a\!\times\!\mathbf{p}_e\right)\!\cdot\!\mathbf{s}_e,
\end{array}
\end{equation}
with implicit summation over $a=1,2$. We have omitted retardation terms, which were considered in~\cite{Korobov20} and shown to be negligibly small. The second-order contribution arises from various terms of the Breit-Pauli Hamiltonian:
\begin{equation} \label{so6-2nd-order}
\begin{array}{@{}l}\displaystyle
\Delta E_{so(6)}^{2^{nd}-order} = \Delta E_{so\hbox{-}H_B} + \Delta E_{so\hbox{-}ret} + \Delta E_{so\hbox{-}so}^{(1)},
\\[3mm]\displaystyle
\Delta E_{so\hbox{-}H_B} =
   2\left\langle
      H_{so} Q (E_0-H_0)^{-1} Q H_B
   \right\rangle,
\\[3mm]\displaystyle
\Delta E_{so\hbox{-}ret} =
   2\left\langle
      H_{so} Q (E_0-H_0)^{-1} Q H_{ret}
   \right\rangle,
\\[3mm]\displaystyle
\Delta E_{so\hbox{-}so}^{(1)} =
   \left\langle
      H_{so} Q (E_0-H_0)^{-1} Q H_{so}
   \right\rangle^{(1)}.
\end{array}
\end{equation}
where $H_0$ and $E_0$ are respectively the nonrelativistic Hamiltonian and energy, and $Q$ is a projection operator on a subspace orthogonal to $\psi_0$. $A^{(k)}$ denotes the term of rank $k$ in the decomposition of $A$ as a sum of irreducible tensor operators. The involved terms of the Breit Pauli-Hamiltonian are
\begin{equation} \label{BP1}
\begin{array}{@{}l}\displaystyle
H_B = -\frac{p_e^4}{8m^3}+\frac{Z_a \pi}{2m^2} \delta(\br_a),
\\[3mm]\displaystyle
H_{ret} =
   \frac{Z_a}{2}\>
      \frac{p_e^i}{m}\left(\frac{\delta^{ij}}{r_a}+\frac{r_1^ir_1^j}{r_a^3}\right)\frac{P_a^j}{M_a},
\\[3mm]\displaystyle
H_{so} =
   \frac{Z_a(1\!+\!2a_e)}{2m^2}\,\frac{(\mathbf{r}_a\!\times\!\mathbf{p}_e)\!\cdot\!\mathbf{s}_e}{r_a^3}
   -\frac{Z_a(1\!+\!a_e)}{mM_a}\,\frac{(\mathbf{r}_a\!\times\!\mathbf{P}_a)\!\cdot\!\mathbf{s}_e}{r_a^3}\,.
\end{array}
\end{equation}
We now turn to the electron-nucleus spin-spin tensor interaction. Similarly, we have
\begin{equation}
\Delta E_{ss(6)}^{(2)} = \langle H_{ss(6)}^{(2)} \rangle + \Delta E_{ss(6)}^{(2)2^{nd}-order},
\end{equation}
where the $m\alpha^6$-order effective Hamiltonian is
\begin{equation}\label{ss6-1st-order}
\begin{array}{@{}l}\displaystyle
H_{ss(6)}^{(2)} = c_W \mathcal{U}_W^{(2)} + c_S \mathcal{U}_{CM}^{(2)},
\\[3mm]\displaystyle
\mathcal{U}_{W}^{(2)} =
   - \frac{1}{4m^2}\,
      \left\{p_e^2,
      \frac{r_a^2\boldsymbol{\mu}_e\!\cdot\!\boldsymbol{\mu}_a
                 \!-\!3(\boldsymbol{\mu}_e\!\cdot\!\mathbf{r}_a)
                       (\boldsymbol{\mu}_a\!\cdot\!\mathbf{r}_a)}{r_a^5}
      \right\}
\\[3mm]\displaystyle
\mathcal{U}_{CM}^{(2)} =
    -\frac{Z_a}{6m}
    \frac{r_a^2\boldsymbol{\mu}_e\!\cdot\!\boldsymbol{\mu}_a
             -3(\boldsymbol{\mu}_e\!\cdot\!\mathbf{r}_a)(\boldsymbol{\mu}_a\!\cdot\!\mathbf{r}_a)}{r_a^6}
\\[3mm]\displaystyle\hspace{10.5mm}
    -\frac{1}{6m}\,
    \left[
       Z_2\frac{(\mathbf{r}_1\!\cdot\!\mathbf{r}_2)\boldsymbol{\mu}_e\!\cdot\!\boldsymbol{\mu}_1
             -3(\boldsymbol{\mu}_e\!\cdot\!\mathbf{r}_1)(\boldsymbol{\mu}_1\!\cdot\!\mathbf{r}_2)}{r_1^3r_2^3}
      +Z_1\frac{(\mathbf{r}_1\!\cdot\!\mathbf{r}_2)\boldsymbol{\mu}_e\!\cdot\!\boldsymbol{\mu}_2
             -3(\boldsymbol{\mu}_e\!\cdot\!\mathbf{r}_2)(\boldsymbol{\mu}_2\!\cdot\!\mathbf{r}_1)}{r_1^3r_2^3}
    \right].
\end{array}
\end{equation}
Here, $\boldsymbol{\mu}_e$ and $\boldsymbol{\mu}_a$ are the electronic and nuclear magnetic moments. Neglecting the electron's anomalous magnetic moment, we get $\boldsymbol{\mu}_e = -(|e|/m) \mathbf{s}_e$. In H$_2^+$, $\boldsymbol{\mu}_a = 2 \mu_p \mu_N \mathbf{I}_a$, where $\mu_p$ is the proton's magnetic moment in units of the nuclear Bohr magneton $\mu_N$, and $\mathbf{I}_a$ the spin operator of nucleus $a$. The second-order contribution is
\begin{equation} \label{ss6-2nd-order}
\begin{array}{@{}l}\displaystyle
\Delta E_{ss(6)}^{(2)2^{nd}-order} = \Delta E_{ss\hbox{-}H_B}^{(2)} + \Delta E_{so\hbox{-}ss}^{(2)} + \Delta E_{so\hbox{-}so_N}^{(2)},
\\[3mm]\displaystyle
\Delta E_{ss\hbox{-}H_B}^{(2)} =
   2\left\langle
      H_{ss}^{(2)} Q (E_0-H_0)^{-1} Q H_B
   \right\rangle,
\\[3mm]\displaystyle
\Delta E_{so\hbox{-}ss}^{(2)} =
   2\left\langle
      H_{ss}^{(2)} Q (E_0-H_0)^{-1} Q H_{so}
   \right\rangle^{(2)},
\\[3mm]\displaystyle
\Delta E_{so\hbox{-}so_N}^{(2)} =
   2\left\langle
      H_{so} Q (E_0-H_0)^{-1} Q H_{so_N}
   \right\rangle^{(2)}.
\end{array}
\end{equation}
It involves two additional terms of the Breit-Pauli Hamiltonian:
\begin{equation} \label{BP2}
\begin{array}{@{}l}\displaystyle
H_{ss}^{(2)} =
   \left[
      \frac{\boldsymbol{\mu}_e\!\cdot\!\boldsymbol{\mu}_a}{r_a^3}
      -3\frac{(\boldsymbol{\mu}_e\!\cdot\!\mathbf{r}_a)(\boldsymbol{\mu}_a\!\cdot\!\mathbf{r}_a)}{r_a^5}
   \right]
   -\frac{8\pi\alpha}{3}\boldsymbol{\mu}_e\!\cdot\!\boldsymbol{\mu}_a\delta(\mathbf{r}_a),
\\[3mm]\displaystyle
H_{so_N} =
   \frac{1}{m}\,\frac{(\mathbf{r}_a\!\times\!\mathbf{p}_e)\!\cdot\!\boldsymbol{\mu}_a}{r_a^3}
   -\frac{1}{M_a}\left(1-\frac{Z_am_pI_a}{M_a\mu_a}\right) \, \frac{(\mathbf{r}_a\!\times\!\mathbf{P}_a)\!\cdot\!\boldsymbol{\mu}_a}{r_a^3}\,.
\end{array}
\end{equation}
We have changed our notations of the first-order terms with respect to Ref.~\cite{Korobov20} in order to clearly identify their link to the terms of the NRQED Hamiltonian. $\mathcal{U}_{CM}$ and $\mathcal{U}_{MM_N}$ denote seagull terms with simultaneous exchange of a Coulomb and a magnetic photon ($CM$), and of two magnetic photons at the nucleus ($MM_N$). The correspondence with notations used in our earlier work~\cite{Korobov20} is the following:
\begin{equation}
\mathcal{U}_W \leftrightarrow \mathcal{U}_{2b} \,, \hspace{3mm} \mathcal{U}_{Y_1} \leftrightarrow \mathcal{U}_{1b} \,, \hspace{3mm} \mathcal{U}_{CM} \leftrightarrow \mathcal{U}_{5a} \,, \hspace{3mm} \mathcal{U}_{MM_N} \leftrightarrow \mathcal{U}_{6b} \,, \hspace{3mm} \mathcal{U}_W^{(2)} \leftrightarrow \mathcal{U}_{2d}^{(2)} \,, \hspace{3mm} \mathcal{U}_{CM}^{(2)} \leftrightarrow \mathcal{U}_{5b}^{(2)} \,.
\end{equation}
None of the coefficients involved in the terms listed in this section have any logarithmic contribution at first order in~$\alpha$ (see Eq.~(\ref{NRQED-coeff})). One can conclude that these terms do not contribute to the order $m\alpha^7 \ln(\alpha)$. Since the nonlogarithmic $m\alpha^7$-order correction is not considered in the present work, in our numerical calculations we truncate the expressions of all coefficients at zero order in~$\alpha$.

\subsection{Terms contributing at the order $m\alpha^7 \ln(\alpha)$}

Contributions at this order stem from spin-dependent coefficients of the NRQED Hamiltonian that depend on $\ln(\alpha)$, i.e. $c_{q^2}$ and $c_{Y_2}$, and can be derived using perturbation theory as done in~\cite{Korobov20,Haidar20}. The first contribution is from a transverse photon exchange with the $c_{q^2}$ term on the electron side and a dipole vertex (labeled 2N in Eq.~(7) of~\cite{Korobov20}) on the nucleus side. The corresponding effective potential in momentum space is
\begin{align}
\mathcal{U}_{q^2} &= \left[ \frac{ie}{8m^3} \bq^2 \left( \bsigma\!\times\!\bq \right) \right]^k \left[ Z_a e  \frac{ (\mathbf{P}_a + \mathbf{P}'_a)}{2 M_a} \right]^l \left[ -\frac{1}{\bq^2} \left( \delta^{kl} - \frac{q^k q^l}{\bq^2} \right) \right] \nonumber \\
&= \frac{i Z_a e^2}{16 m^3 M_a} \left( \bsigma\!\times\!\bq \right)\!\cdot\! \left(\mathbf{P}_a + \mathbf{P}'_a \right) = -\frac{i Z_a e^2}{16 m^3 M_a} \left[ \bq\!\times\!\left(\mathbf{P}_a + \mathbf{P}'_a \right) \right]\!\cdot\!\bsigma \,.
\end{align}
After Fourier transform, the effective potential in real space is found to be
\begin{equation}
\mathcal{U}_{q^2} = \frac{i Z_a e^2}{8 m^3 M_a} \left( \bp_e \!\times\! 4\pi\delta(\mathbf{r}_a) \mathbf{P}_a - \mathbf{P}_a \!\times\! 4\pi\delta(\mathbf{r}_a) \bp_e \right) \!\cdot\! \bs_e \,.
\end{equation}
The other contribution is due to a Coulomb photon exchange with the $c_{Y_2}$ term on the electron side and a Coulomb vertex (2N in Eq.~(7) of~\cite{Korobov20}) on the nucleus side:
\begin{equation}
\mathcal{U}_{Y_2} = \left[ \frac{ie}{8m^4} \bsigma\!\cdot\!\left[ \bq\!\times\! (\bp'+\bp) \right] \bq^2 \right] [-Z_a e] \left[ \frac{1}{\bq^2} \right] = -\frac {i Z_a e^2}{4 m^4} \left( \bq\!\times\!\bp \right)\!\cdot\!\bsigma \,,
\end{equation}
which yields for the real-space effective potential:
\begin{equation}
\mathcal{U}_{Y_2} = \frac{i Z_a e^2}{2 m^4} \left( \bp_e \!\times\! 4\pi\delta(\mathbf{r}_a) \bp_e \right) \!\cdot\! \bs_e \,.
\end{equation}
Both terms contribute to the spin-orbit interaction. The total effective potential of order $m\alpha^7 \ln(\alpha)$ is thus obtained as:
\begin{equation} \label{so7-1st-order}
H_{so(7\ln)} = c_{q^2} \mathcal{U}_{q^2} + c_{Y_2} \mathcal{U}_{Y_2} \,,
\end{equation}
with
\begin{equation}
c_{q^2} \equiv \frac{4}{3} \frac{\alpha}{\pi} \ln \left( \alpha^{-2} \right) \,, \hspace{3mm} c_{Y_2} \equiv -\frac{1}{3} \frac{\alpha}{\pi} \ln \left( \alpha^{-2} \right) .
\end{equation}
Note that the nonrecoil term $\mathcal{U}_{q^2}$ had been obtained for an electron in an external potential in~\cite{Jentschura05} (see also~\cite{Pachucki99}), but the recoil term $\mathcal{U}_{Y_2}$ had not been considered so far, to the best of our knowledge. There is also a second-order perturbation term:
\begin{equation}\label{so7-2nd-order}
\Delta E_{so(7\ln)}^{2^{nd}-order} = 2 \; \langle H_{so} Q (E_0 - H_0)^{-1} Q H_{(5\ln)} \rangle \,,
\end{equation}
where
\begin{equation}
\label{Hprime5}
H_{(5\ln)} = \alpha^3 \frac{4}{3} \ln(\alpha^{-2}) Z_a \delta(\br_a)
\end{equation}
is the logarithmic part of the effective Hamiltonian describing leading-order radiative corrections. The total correction to the spin-orbit interaction at this order is
\begin{equation}
\Delta E_{so(7\ln)} = \langle H_{so(7\ln)} \rangle + \Delta E_{so(7\ln)}^{2^{nd}-order}.
\end{equation}
From the above discussion of logarithmic terms in the NRQED Hamiltonian coefficients, it is clear that there are no effective potentials contributing to the spin-spin tensor interaction at the order $m\alpha^7\ln(\alpha)$. The only contribution is thus the second-order term
\begin{equation}\label{ss7-2nd-order}
\Delta E_{ss(7\ln)}^{(2)} =   2 \; \langle H_{ss}^{(2)} Q (E_0 - H_0)^{-1} Q H_{(5\ln)} \rangle.
\end{equation}
The explicit expressions of corrections to the spin-orbit and spin-spin tensor coefficients, which in the H$_2^+$ case are denoted by $c_e$ and $d_1$ respectively (see Eq.~(3) of~\cite{Korobov06} for definitions), in terms of reduced matrix elements of the effective potentials listed in this section, are given in Appendix~\ref{delta-ce-d1} (see~\cite{Haidar-thesis} for details).

\section{Numerical results}

\subsection{Variational method}

The main features of our numerical method have been described in Ref.~\cite{Korobov20}. The wave function for a rovibrational state $(v,L)$ is expanded in terms of exponentials of interparticle distances in the following way:
\begin{equation}\label{var_expansion}
\begin{array}{@{}l}
\displaystyle \Psi_0 (\mathbf{R},\mathbf{r}_1) =
       \sum_{l_1+l_2=L}
         \mathcal{Y}^{l_1l_2}_{LM}(\hat{\mathbf{R}},\hat{\mathbf{r}}_1)
         G_{l_1l_2}(R,r_1,r_2),
\\[5mm]\displaystyle
\mathcal{Y}^{l_1l_2}_{LM}(\hat{\mathbf{R}},\hat{\mathbf{r}}_1) = R^{l_1} r_1^{l_2} \left\{ Y_{l_1}(\hat{\mathbf{R}}) \otimes Y_{l_2}(\hat{\mathbf{r}}_1) \right\}_{LM},
\\[1mm]\displaystyle
G_{l_1l_2}(R,r_1,r_2) = \sum_{n=1}^{N/2} \Big\{C_n\,\mbox{Re} \bigl[e^{-\alpha_n R-\beta_n r_1-\gamma_n r_2}\bigr]
+D_n\,\mbox{Im} \bigl[e^{-\alpha_n R-\beta_n r_1-\gamma_n r_2}\bigr] \Big\}.
\end{array}
\end{equation}
The complex exponents $\alpha_n$, $\beta_n$, $\gamma_n$ are generated in a pseudorandom way in several intervals, which play the role of variational parameters. We have used 2 intervals for the lower vibrational states ($0 \leq v \leq 4$) and 4 for higher states ($5 \leq v \leq 9$).

\subsection{Second-order terms}

Second-order terms have a general expression of the type $\langle A Q (E_0 - H_0)^{-1} Q B \rangle$. They are evaluated by solving numerically the equation
\begin{equation} \label{psi1}
(E_0 - H_0) \psi^{(1)} = (B - \langle B \rangle) \, \psi_0,
\end{equation}
and calculating the scalar product $\langle \Psi_0 | A | \psi^{(1)} \rangle$. In order to solve Eq.~(\ref{psi1}), $\psi^{(1)}$ is expanded in an ``intermediate'' variational basis following Eq.~(\ref{var_expansion}). As discussed in~\cite{Korobov20}, the most difficult contributions for numerical evaluation are the singular second-order terms: $\Delta E_{so\hbox{-}H_B}$ [Eq.~(\ref{so6-2nd-order})], $\Delta E_{ss\hbox{-}H_B}$, [Eq.~(\ref{ss6-2nd-order})], $\Delta E_{so(7\ln)}^{2^{nd}-order}$ [Eq.~(\ref{so7-2nd-order})], and $\Delta E_{ss(7\ln)}^{(2)}$ [Eq.~(\ref{ss7-2nd-order})]. Indeed, if $B = H_B$ or $B = H_{(5\ln)}$ in Eq.~(\ref{psi1}), the intermediate wave function $\psi^{(1)}$ behaves like $1/r_1$ ($1/r_2$) at small electron-nucleus distances, resulting in very slow convergence. To circumvent this problem, we rewrite the second-order energy shift as~\cite{Korobov20}
\begin{equation}
\left\langle A Q (E_0-H_0)^{-1} Q B \right\rangle = \left\langle A Q (E_0-H_0)^{-1} Q B' \right\rangle + \langle U A \rangle - \langle U \rangle \langle A \rangle,
\end{equation}
where
\begin{equation} \label{sing-sep}
\begin{array}{@{}l}
\displaystyle
U = \frac{c_1}{r_1} + \frac{c_2}{r_2},
\\[3mm]\displaystyle
B' = B - (E_0 - H_0)U - U(E_0 - H_0).
\end{array}
\end{equation}
For the case $B = H_B$, we have
\begin{equation}
c_a = \frac{\mu_a (2 \mu_a - m_e)}{4 m_e^3} Z_a ,
\end{equation}
with $\mu_a = M_a m_e / (M_a + m_e)$, and for $B = H_{(5\ln)}$,
\begin{equation}
c_a = \alpha^3 \frac{4}{3} \ln(\alpha^{-2}) Z_a \times \left( -\frac{\mu_a Z_a}{2 \pi} \right).
\end{equation}
The replacement of $B$ by $B'$ in Eq.~(\ref{psi1}) reduces the singularity of the intermediate wavefunction. The remaining logarithmic singularity $\psi^{(1)} \sim \ln(r_1)$ ($\ln(r_2)$) still slows down the convergence, and necessitates expanding $\psi^{(1)}$ in a ``multilayer'' basis set (see Table I in~\cite{Korobov20} for an example), where the first subsets (between 2 and 4) approximate the regular part, and 8 additional subsets contain growing exponents $\beta_n$ ($\gamma_n$) up to 10$^4$ in order to reproduce the singular behavior.

\subsection{Convergence study}

We now analyze the convergence of our numerical results. For first-order terms, sufficient accuracy is quite easily obtained; for illustration, the reduced matrix elements involved in calculation of $\mathcal{U}_W$ and $\mathcal{U}_{Y_1}$ [Eq.~(\ref{so6-1st-order})] are shown in Table~\ref{conv-1st-order}. Convergence is slower for the terms involving $(\mathbf{r}_a\!\times\!\mathbf{p}_e)$, which are related to the electronic contribution to the total orbital momentum, because their nonzero value entirely comes from the smaller ``non-$\sigma$'' (i.e. $l_2 \neq 0$ in Eq.~(\ref{var_expansion})) components of the wave function. For the same reason, these matrix elements are smaller than those involving $(\mathbf{r}_a\!\times\!\mathbf{P}_a)$ by a factor of order $m/M_a \sim 10^{-3}$. Overall, first-order terms are obtained with at least 3-4 significant digits of accuracy.
\begin{table}[h]
\begin{tabular}{lllll}
\hline\hline
$N$ & $p_e^2 \frac{1}{r_1^3}[\mathbf{r}_1\!\times\!\mathbf{p}_e]$ & $p_e^2 \frac{1}{r_2^3}[\mathbf{r}_2\!\times\!\mathbf{p}_e]$ & $p_e^2 \frac{1}{r_1^3}[\mathbf{r}_1\!\times\!\mathbf{P}_1]$ & $p_e^2 \frac{1}{r_2^3}[\mathbf{r}_2\!\times\!\mathbf{P}_2]$ \\
\hline
1400 & -0.209756[-03] & -0.211145[-03] & -0.718198 & -0.718358 \\
1600 & -0.211462[-03] & -0.212048[-03] & -0.718194 & -0.718138 \\
1800 & -0.210752[-03] & -0.210806[-03] & -0.718136 & -0.718143 \\
2000 & -0.210069[-03] & -0.211858[-03] & -0.718145 & -0.718142 \\
2200 & -0.210909[-03] & -0.210218[-03] &           &           \\
2400 & -0.211099[-03] & -0.211191[-03] &           &           \\
2600 & -0.211024[-03] & -0.211042[-03] &           &           \\
\hline\hline
\end{tabular}
\caption{Convergence of the reduced matrix elements involved in the first-order terms $\mathcal{U}_W$ and $\mathcal{U}_{Y_1}$ [Eq.~(\ref{so6-1st-order})] for the ($L=1,v=4$) state of H$_2^+$ (values are given in a.u). \label{conv-1st-order}}
\end{table}

Second-order terms, especially the singular terms discussed above, require heavier numerical calculations. This is illustrated in Table~\ref{conv-2nd-order}, which shows the convergence of $\Delta E_{so\hbox{-}H_B}$ [Eq.~(\ref{so6-2nd-order})]. The quantities appearing in this Table are
\begin{equation}
A_a = \left \langle vL \left \| \frac{1}{r_a^3}(\mathbf{r}_{a}\times\mathbf{p}_e) Q (E_0-H_0)^{-1} Q H'_B  \right \| vL \right  \rangle,
\end{equation}
where $H'_B$ is the effective Hamiltonian obtained by applying the transformation~(\ref{sing-sep}) to $B = H_B$, whereas the left-hand side appears in the nonrecoil part of $H_{so}$ [Eq.~(\ref{BP1})]. From Table~\ref{conv-2nd-order} it can be estimated that these matrix elements are obtained with 3 significant digits. Second-order matrix elements involving $(\mathbf{r}_{a}\times\mathbf{P}_a)$ in the left-hand side, corresponding to the recoil part of $H_{so}$, exhibit faster convergence (not shown in Table~\ref{conv-2nd-order}), similarly to what was discussed for first-order terms.

\begin{table}[h]
\begin{tabular}{llllll}
\hline\hline
$N$ & $A_1$ & $A_2$ & $a^e_0$ & $a^e_+$ & $\displaystyle \Delta c_e^{(6)} |_{so^e-so^e}$ \\
\hline
8000  & -0.746134[-04] & -0.801218[-04] & -0.12654393[-01] & -0.12680066[-01] & -0.6418[-05] \\
10000 & -0.795165[-04] & -0.797075[-04] & -0.12657847[-01] & -0.12680231[-01] & -0.5596[-05] \\
12000 & -0.796812[-04] & -0.797663[-04] & -0.12657987[-01] & -0.12680252[-01] & -0.5566[-05] \\
14000 & -0.796931[-04] & -0.797285[-04] & -0.12658040[-01] & -0.12680278[-01] & -0.5560[-05] \\
16000 & -0.797234[-04] & -0.797646[-04] & -0.12658073[-01] & -0.12680294[-01] & -0.5555[-05] \\
\hline\hline
\end{tabular}
\caption{Convergence of second-order terms contributing to $\Delta E_{so\hbox{-}H_B}$ and to $\displaystyle\Delta E_{so\hbox{-}so}^{(1)}$ for the ($L=1,v=4$) state of H$_2^+$ (values are given in a.u). \label{conv-2nd-order}}
\end{table}
A term that deserves a separate discussion, $\displaystyle\Delta E_{so\hbox{-}so}^{(1)}$ [Eq.~(\ref{so6-2nd-order})], is also shown in Table~\ref{conv-2nd-order}. Again, only the contributions from the nonrecoil part of $H_{so}$, which are the most difficult to converge, are shown. These contributions, denoted by $a^e_0$ and $a^e_+$, are obtained from Eq.~(\ref{components-so-so}) by replacing $\mathbf{A_{so}}$ with $\mathbf{A^e_{so}}$, which only includes the first term of $H_{so}$:
\begin{equation}
\mathbf{A^e_{so}} = \frac{Z_a}{2m^2}\,\frac{(\mathbf{r}_a\!\times\!\mathbf{p}_e)}{r_a^3}
\end{equation}
The corresponding contribution to $c_e$ is (see Eq.~(\ref{deltace-so-so}))
\begin{equation}
\Delta c_e^{(6)} |_{so^e-so^e} = -\frac{1}{2} \frac{1}{L(L+1)} \biggl[ (L+1)a^e_- + a^e_0 -L a^e_+ \biggr].
\end{equation}
As can be seen from Table~\ref{conv-2nd-order}, the quantities $a^e_0$, $a^e_+$ converge more rapidly than $A_1$ and $A_2$, in accordance with the fact that $H_{so}$ is less singular than $H'_B$. However, due to a quasi cancellation between the different angular momentum components, they are larger than the total contribution $\Delta c_e^{(6)} |_{so-so}$ by several orders of magnitude. As a consequence, they need to be calculated with a high relative accuracy, which requires using a large variational basis. From the results of Table~\ref{conv-2nd-order}, the numerical uncertainty of $\Delta c_e^{(6)} |_{so-so}$ may be conservatively estimated to $10^{-7} E_h \alpha^4$ (where $E_h$ is the Hartree energy), i.e. less than 2~Hz.

\subsection{Results}

The values of all the contributions to the spin-orbit coefficient $c_e$ are given in Table~\ref{cetable-h2plus} for a few states of interest for experiments. Note that the term $\mathcal{U}_{MM_N}$ [Eq.~(\ref{so6-1st-order})] was omitted, because it was found to be smaller than 1~Hz, which is negligible with respect to the overall uncertainty. Our new theoretical values of $c_e$ can be found in the last column. Complete results for the rovibrational states $(0 \leq L \leq 4, 0 \leq v \leq 9)$ are given in the Appendix~\ref{app-results}.

The numerical uncertainty is dominated by the singular second-order term $\Delta E_{so\hbox{-}H_B}$; from the convergence study shown in the previous paragraph and similar tests performed for higher vibrational states, it is estimated to be smaller than 10~Hz for all rovibrational states. The theoretical uncertainty is mainly due to the yet uncalculated nonlogarithmic correction of order $m\alpha^7$~\cite{Pachucki99,Pachucki09}. We estimate it to about one third of the $m\alpha^7\ln(\alpha)$ correction, which corresponds to 100-150~Hz or 3-4~ppm.
\begin{table} [h!]
\small
\begin{tabular}{|@{\hspace{1mm}}c@{\hspace{1mm}}|@{\hspace{1mm}}c@{\hspace{1mm}}|@{\hspace{1mm}}c@{\hspace{1mm}}|@{\hspace{1mm}}c@{\hspace{1mm}}|@{\hspace{1mm}}c@{\hspace{1mm}}
|@{\hspace{1mm}}c@{\hspace{1mm}}|@{\hspace{1mm}}c@{\hspace{1mm}}|@{\hspace{1mm}}c@{\hspace{1mm}}|@{\hspace{1mm}}c@{\hspace{1mm}}
|@{\hspace{1mm}}c@{\hspace{1mm}}|@{\hspace{1mm}}c@{\hspace{1mm}}|@{\hspace{1mm}}c@{\hspace{1mm}}|@{\hspace{1mm}}c@{\hspace{1mm}}|@{\hspace{1mm}}c@{\hspace{1mm}}|}
\hline
$(L,v)$ &$\displaystyle  c_e^{(BP)}$& $\displaystyle\mathcal{U}_{Y_1}$ & $\displaystyle\mathcal{U}_{W}$ & $\displaystyle\mathcal{U}_{CM}$ & $\displaystyle\Delta E_{so\hbox{-}H_B}$ &$\displaystyle\Delta E_{so\hbox{-}so}^{(1)}$& $\displaystyle\Delta E_{so\hbox{-}ret}$ & $\displaystyle
\Delta  c_e^{(6)}$ & $\displaystyle\mathcal{U}_{Y_1}$ & $\displaystyle\mathcal{U}_{q^2}$ & $\displaystyle \Delta E_{so\hbox{-}H_{(5\ln)}}$ & $\displaystyle
\Delta  c_e^{(7\ln)}$ & $\displaystyle c_e$ (this work) \\[2mm]
\hline
(1,0) & 42\,416.318 & 1.551 & -3.631 & 0.028 & 2.765 & 0.414 & 0.333 & 1.460 & -0.035 & 0.060 & -0.486 & -0.460 & 42\,417.32(15) \\
(1,4) & 32\,654.638 & 1.205 & -2.979 & 0.055 & 2.154 & 0.325 & 0.261 & 1.020 & -0.027 & 0.049 & -0.364 & -0.342 & 32\,655.32(11) \\
(1,5) & 30\,437.196 & 1.127 & -2.813 & 0.058 & 2.010 & 0.305 & 0.239 & 0.925 & -0.025 & 0.046 & -0.337 & -0.316 & 30\,437.80(11) \\
(1,6) & 28\,280.421 & 1.049 & -2.645 & 0.059 & 1.858 & 0.283 & 0.220 & 0.824 & -0.023 & 0.044 & -0.312 & -0.292 & 28\,280.95(10) \\
(2,0) & 42\,162.530 & 1.542 & -3.601 & 0.027 & 2.733 & 0.412 & 0.336 & 1.447 & -0.034 & 0.060 & -0.481 & -0.456 & 42\,163.52(15) \\
(2,1) & 39\,571.598 & 1.451 & -3.440 & 0.036 & 2.579 & 0.388 & 0.311 & 1.326 & -0.032 & 0.057 & -0.448 & -0.424 & 39\,572.50(14) \\
\hline
\end{tabular}
\caption{\label{cetable-h2plus} Corrections to the spin-orbit interaction coefficient $c_e$ for a few rovibrational states of H$_2^+$ (in kHz). The leading-order (Breit-Pauli) value $c_e^{(BP)}$ (Ref.~\cite{Korobov06}) is given in column 2. Columns 3-5 and 6-8 are respectively the first-order and second-order contributions [Eqs.~(\ref{so6-1st-order}) and (\ref{so6-2nd-order})] at the  m$\alpha^6$ order, and the total correction at this order,  $\Delta  c_e^{(6)}$, is given in column 9. Columns 10-12 are the first-order [Eq.~(\ref{so7-1st-order})] and second-order [Eq.~(\ref{so7-2nd-order})] contributions at the m$\alpha^7\ln(\alpha)$ order, respectively. The total correction at this order, $\Delta  c_e^{(7\ln)}$, is given in column 13. The last column is our new value of $c_e$. Its estimated uncertainty (equal to one third of $\Delta  c_e^{(7\ln)}$) is indicated between parentheses.}
\end{table}

Regarding the spin-spin tensor interactions, we write the related term of the H$_2^+$ effective spin Hamiltonian~\cite{Korobov06} in the following way:
\begin{equation}
H_{\rm eff}^{ss(2)} = d_1 \left( 2 \bL^2 (\bs_e\!\cdot\!\bI) - 3 \left[ (\bL\!\cdot\!\bs_e)(\bL\!\cdot\!\bI) + (\bL\!\cdot\!\bI)(\bL\!\cdot\!\bs_e) \right] \right)
\end{equation}
This definition differs from that of Ref.~\cite{Korobov06} by a factor $3(2L-1)(2L+3)=15$ (for $L=1$), but coincides with that of the $E_6$ coefficient in the HD$^+$ effective spin Hamiltonian~\cite{Bakalov06}, which facilitates future comparison between H$_2^+$ and HD$^+$. The values of all the contributions to the $d_1$ coefficient are given in Table~\ref{d1results} for a few $L=1$ states, whereas complete results for the ro-vibrational states $(0 \leq L \leq 4, 0 \leq v \leq 9)$ are given in the Appendix~\ref{app-results}. The second-order terms $\Delta E_{so\hbox{-}ss}^{(2)}$ and $\Delta E_{so\hbox{-}so_N}^{(2)}$ have been omitted, because they were found to be much smaller than the overall uncertainty. The numerical uncertainty, dominated by the singular second-order term $\Delta E_{ss\hbox{-}H_B}^{(2)}$, is estimated to be smaller than 1~Hz for all rovibrational states. Similarly to the spin-orbit coefficient, the theoretical uncertainty due to the yet uncalculated nonlogarithmic correction of order $m\alpha^7$ is estimated to about one third of the $m\alpha^7\ln(\alpha)$ correction, which corresponds to 10-20~Hz or about 2~ppm.

\begin{table} [h!]
\small
\begin{tabular}{|@{\hspace{1mm}}c@{\hspace{1mm}}|@{\hspace{1mm}}c@{\hspace{1mm}}|@{\hspace{1mm}}c@{\hspace{1mm}}|@{\hspace{1mm}}c@{\hspace{1mm}}
|@{\hspace{1mm}}c@{\hspace{1mm}}|@{\hspace{1mm}}c@{\hspace{1mm}}|@{\hspace{1mm}}c@{\hspace{1mm}}|@{\hspace{1mm}}c@{\hspace{1mm}}|}
\hline
$(L,v)$ & $\displaystyle d_1^{(BP)}$ & $\displaystyle\mathcal{U}_W^{(2)}$&$\displaystyle\mathcal{U}_{CM}^{(2)}$ & $\displaystyle\Delta E_{ss\hbox{-}H_B}^{(2)}$ & $\displaystyle\Delta d_1^{(6)}$ &  $\displaystyle\Delta d_1^{(7\ln)}$ & $\displaystyle d_1$ (this work) \\[2mm]
\hline
(1,0) & 8\,565.983 & -0.802 & 0.092 & 0.951 & 0.241 & -0.050 &  8\,566.174(17) \\
(1,4) & 6\,537.247 & -0.642 & 0.079 & 0.740 & 0.178 & -0.039 &  6\,537.386(13) \\
(1,5) & 6\,080.287 & -0.603 & 0.076 & 0.676 & 0.149 & -0.036 &  6\,080.400(12) \\
(1,6) & 5\,637.524 & -0.564 & 0.072 & 0.629 & 0.137 & -0.033 &  5\,637.627(11) \\
\hline
\end{tabular}
\caption{Corrections to the spin-spin tensor interaction coefficient $d_1$ for a few rovibrational states of H$_2^+$ (in kHz). The leading-order (Breit-Pauli) value $\displaystyle d_1^{(BP)}$ (Ref.~\cite{Korobov06}) is given in column 2. Columns 3-4 and 5 are respectively the first-order and second-order contributions [Eqs. (\ref{ss6-1st-order}) and (\ref{ss6-2nd-order})] at the $m\alpha^6$ order. The total correction at this order,  $\displaystyle \Delta  d_1^{(6)}$, is given in column 6. Column 7 is the second-order contribution at the $m\alpha^7\ln(\alpha)$ order [Eq.~(\ref{ss7-2nd-order})]. The last column is our new value for $d_1$. Its estimated uncertainty (equal to one third of $\Delta  d_1^{(7\ln)}$) is indicated between parentheses. To match the notations of Ref.~\cite{Korobov06}, all values should be multiplied by $3(2L-1)(2L+3)=15$.\label{d1results}}
\end{table}

\section{Comparison with experiments}

We now use our new values of the $c_e$ and $d_1$ coefficients to obtain improved theoretical predictions of the hyperfine intervals measured in~\cite{Menasian73}. To do this, we diagonalize the effective spin Hamiltonian of Ref.~\cite{Korobov06}. The values of the spin-spin contact interaction coefficient $b_F$ are taken from~\cite{Karr20}; it is worth recalling that they have been found to be in excellent agreement with experimental rf spectroscopy data~\cite{Jefferts69}. The smaller hyperfine coefficients $c_I$ and $d_2$, which respectively describe the nuclear spin-orbit and the proton-proton spin-spin tensor interaction, are calculated in the framework of the Breit-Hamiltonian with account of the electron's anomalous magnetic moment~\cite{Korobov06}. The values of all the coefficients used here can be found in the Appendix~\ref{hfs-coeff}.

In order to estimate the uncertainties of the theoretical hyperfine intervals $f_v$, we calculated the derivatives
\begin{equation}
\gamma_{c_e,v} = \frac{\partial f_v}{\partial c_e} \, , \;\; \gamma_{c_I,v} = \frac{\partial f_v}{\partial c_I} \, , \;\; \ldots
\end{equation}
Their values for the three rovibrational levels of interest are given in Appendix~\ref{derivatives}. The uncertainty of $f_v$ is calculated via propagation of the uncertainties of the hyperfine coefficients. Note that this uncertainty only weakly depends on our assumptions regarding correlations, because it is dominated by the uncertainty of the $c_e$ coefficient, whereas the second largest uncertainty, from $d_1$, is smaller by more than one order of magnitude. Assuming no correlations between uncertainties of different coefficients, the total uncertainty is
\begin{equation}
u(f_v) = \sqrt{ \left( \gamma_{c_e,v} u(c_e,v) \right)^2 + \left( \gamma_{c_I,v} u(c_I,v) \right)^2 +\left( \gamma_{b_F,v} u(b_F,v) \right)^2 + \left( \gamma_{d_1,v} u(d_1,v) \right)^2 + \left( \gamma_{d_2,v} u(d_2,v) \right)^2 }
\end{equation}
The uncertainties $u(c_e)$ and $u(d_1)$ have been estimated above, $u(b_F)$ is taken from~\cite{Karr20}, and for the coefficients calculated at the Breit-Pauli level we take $u(c_I)= \alpha^2 c_I$ and $u(d_2)=\alpha^2 d_2$.

\begin{table}[h!]
\begin{tabular}{ |p{1cm}|p{3cm}|p{4cm}|p{3cm}|}
\hline
$(L,v)$ & Theory \cite{Korobov06}  &Theory (this work)&Experiment~\cite{Menasian73}\\
\hline
(1,4) & 15.371\,0(9) & 15.371\,316(56) & 15.371\,407(2) \\
(1,5) & 14.381\,2(8) & 14.381\,453(52) & 14.381\,513(2) \\
(1,6) & 13.413\,2(7) & 13.413\,397(48) & 13.413\,460(2) \\
\hline
\end{tabular}
\caption{Comparison between theory and experiment for the hyperfine splitting between the $(F=1/2,J=3/2)$ and $(F=1/2,J=1/2)$ states (in MHz). The second column gives the theoretical prediction obtained from calculation of the hyperfine coefficients at the Breit-Pauli level, and the third one is our new prediction including higher-order corrections to $b_F$, $c_e$, and $d_1$. The experimental values are shown in the last column. \label{exp-comparison}}
\end{table}

The comparison between theory and experiment, presented in Table~\ref{exp-comparison}, reveals a reasonable agreement. The observed deviations, which range between 1.2 and 1.6~$\sigma$, may for example be caused by a slight underestimate of the nonlogarithmic correction of order $m\alpha^7$ to the spin-orbit coefficient $c_e$.

In conclusion, we have advanced the hyperfine structure theory in hydrogen molecular ions by calculating higher-order corrections to the spin-orbit and spin-spin tensor interactions. This allowed us to improve the accuracy of the related hyperfine coefficients in H$_2^+$ by about one order of magnitude and reach agreement with rf spectroscopy data at a level of 4-6~ppm. In the future, the theory can be improved further by calculating nonlogarithmic $m\alpha^7$-order corrections to the spin-orbit coefficient. Application to HD$^+$, which has been a subject of several high-precision experiments in recent years, will be presented in a forthcoming paper.

\appendix

\section{Expressions of corrections to the hyperfine coefficients} \label{delta-ce-d1}

All the first-order terms contributing to the spin-orbit interaction, Eqs.~(\ref{so6-1st-order}) and (\ref{so7-1st-order}), as well as the second-order terms $\Delta E_{so-H_B}$, $\Delta E_{so-ret}$ in Eq.~(\ref{so6-2nd-order}) and $\Delta E_{so(7\ln)}^{2^{nd}-order}$ [Eq.~(\ref{so7-2nd-order})], can be written in the form $\langle \mathcal{U}_i \rangle = \langle \mathbf{A}_i \!\cdot\! \mathbf{s_e} \rangle$, where $\mathbf{A}_i$ is a vector operator acting on space variables. The corresponding correction to the spin-orbit coefficient (denoted by $c_e$ in H$_2^+$~\cite{Korobov06}) is then obtained from the Wigner-Eckart theorem as
\begin{equation}
\Delta c_e (v,L) = \frac {\langle v L || \mathbf{A}_i || v L \rangle} { \langle L || \mathbf{L} || L \rangle } = \frac {\langle v L || \mathbf{A}_i || v L \rangle} {\sqrt{L(L+1)(2L+1)} }
\end{equation}
Similarly, the first-order terms contributing to the spin-spin tensor interaction [Eq.~(\ref{ss6-1st-order})], and the second-order terms $\Delta E_{ss\hbox{-}H_B}^{(2)}$ in Eq.~(\ref{ss6-2nd-order}) and $\Delta E_{ss(7\ln)}^{(2)}$ [Eq.~(\ref{ss7-2nd-order})], can be written in the form $\langle \mathcal{U}_i \rangle = \langle \mathbf{T}_i^{(2)} \!\cdot\! \mathbf{U}^{(2)} \rangle$, where $\mathbf{T}_i^{(2)}$ is an operator of rank 2 acting on space variables, and (see Appendix B in~\cite{Korobov20})
\begin{equation}
\mathbf{U}^{(2)}_{\mu} = \{\mathbf{s}_e\otimes\mathbf{I}\}^{(2)}_{\mu}
 = \sqrt{\frac{3}{2}}
   \left[\frac{1}{2}(s_e^iI^j+s_e^jI^j)-\frac{\delta^{ij}}{3}(\mathbf{s}_e\!\cdot\!\mathbf{I})\right]^{(2)}_{\mu}.
\end{equation}
Here, $\mathbf{I} = \mathbf{I}_1  +\mathbf{I}_2$ is the total nuclear spin. Using again the Wigner-Eckart theorem and the relationship
\begin{equation}
(\mathbf{L} \otimes \mathbf{L})^{(2)} \cdot (\mathbf{s}_e \otimes \mathbf{I})^{(2)} =
\frac{1}{2}\sqrt{\frac{3}{2}} \left[(\bL\!\cdot\!\bs_e)(\bL\!\cdot\!\bI)+(\bL\!\cdot\!\bI)(\bL\!\cdot\!\bs_e) - \frac{2}{3} \bL^2 (\bs_e\!\cdot\!\bI) \right] ,
\end{equation}
one gets for the correction to the tensor coefficient (denoted by $d_1$ in H$_2^+$~\cite{Korobov06}):
\begin{equation}
\Delta d_1 (v,L) = -\frac{1}{2\sqrt{6}} \frac {\langle v L || \mathbf{T}_i^{(2)} || v L \rangle} { \langle L || (\mathbf{L}\otimes\mathbf{L})^{(2)} || L \rangle } = - \frac {\langle v L || \mathbf{T}_i^{(2)} || v L \rangle} {2 \sqrt{L(L+1)(2L-1)(2L+1)(2L+3)}}.
\end{equation}
Some of the second-order terms are more complicated because they involve a coupling of two spatial operators of rank 1 or 2. This case was treated in detail in the Appendix B of~\cite{Korobov20}; we only give here the final formula for the term $\Delta E_{so-so}^{(1)}$ in Eq.~(\ref{so6-2nd-order}), as obtained by applying Eqs. (B3) and (B6) of that reference:
\begin{equation} \label{deltace-so-so}
\Delta c_e (v,L) = -\frac{1}{2} \frac{1}{L(L+1)}  \biggl[(L+1)a_- + a_0 -L a_+   \biggr],
\end{equation}
where
\begin{equation} \label{components-so-so}
\begin{array}{@{}l}\displaystyle
a_- = -\frac{1}{2L+1}
   \sum_{n\ne0} \frac{\left\langle vL\|\mathbf{A}_{so}\|v_nL-1\right\rangle
   \left\langle v_nL-1\|\mathbf{A}_{so}\|vL-1\right\rangle}{E_0-E_n}\,,
\\[3mm]\displaystyle
a_0 = \frac{1}{2L+1}
   \sum_{n\ne0} \frac{\left\langle vL\|\mathbf{A}_{so}\|v_nL\right\rangle
   \left\langle v_nL\|\mathbf{A}_{so}\|vL\right\rangle}{E_0-E_n}\,,
\\[3mm]\displaystyle
a_+ = -\frac{1}{2L+1}
   \sum_{n\ne0} \frac{\left\langle vL\|\mathbf{A}_{so}\|v_nL+1\right\rangle
   \left\langle v_nL+1\|\mathbf{A}_{so}\|vL+1\right\rangle}{E_0-E_n}\,.
\end{array}
\end{equation}
$\mathbf{A}_{so}$ is the spatial part of the spin-orbit Hamiltonian $H_{so}$ in Eq.~(\ref{BP1}), i.e. $H_{so} = \mathbf{A}_{so} \!\cdot\! \bs_e$.

\section{Numerical results} \label{app-results}

\begin{table} [h!]
\small
\begin{tabular}{|@{\hspace{1mm}}c@{\hspace{1mm}}|@{\hspace{1mm}}c@{\hspace{1mm}}|@{\hspace{1mm}}c@{\hspace{1mm}}|@{\hspace{1mm}}c@{\hspace{1mm}}|@{\hspace{1mm}}c@{\hspace{1mm}}
|@{\hspace{1mm}}c@{\hspace{1mm}}|@{\hspace{1mm}}c@{\hspace{1mm}}|@{\hspace{1mm}}c@{\hspace{1mm}}|@{\hspace{1mm}}c@{\hspace{1mm}}
|@{\hspace{1mm}}c@{\hspace{1mm}}|@{\hspace{1mm}}c@{\hspace{1mm}}|@{\hspace{1mm}}c@{\hspace{1mm}}|@{\hspace{1mm}}c@{\hspace{1mm}}|@{\hspace{1mm}}c@{\hspace{1mm}}|}
\hline
$(L,v)$ &$\displaystyle c_e^{(BP)}$& $\displaystyle\mathcal{U}_{Y_1}$ & $\displaystyle\mathcal{U}_{W}$ & $\displaystyle\mathcal{U}_{CM}$ & $\displaystyle\Delta E_{so\hbox{-}H_B}$ &$\displaystyle\Delta E_{so\hbox{-}so}^{(1)}$& $\displaystyle\Delta E_{so\hbox{-}ret}$ & $\displaystyle
\Delta  c_e^{(6)}$ & $\displaystyle\mathcal{U}_{Y_2}$ & $\displaystyle\mathcal{U}_{q^2}$ & $\displaystyle \Delta E_{so\hbox{-}H_{(5)}}^{\ln}$ & $\displaystyle
\Delta  c_e^{(7\ln)}$ & $\displaystyle c_e$ (this work) \\[2mm]
\hline
(1,0) & 42\,416.318 & 1.551 & -3.631 & 0.028 & 2.765 & 0.414 & 0.333 & 1.460 & -0.035 & 0.060 & -0.486 & -0.460 & 42\,417.32(15) \\
(1,1) & 39\,812.244 & 1.460 & -3.469 & 0.037 & 2.609 & 0.391 & 0.307 & 1.335 & -0.033 & 0.058 & -0.453 & -0.428 & 39\,813.15(14) \\
(1,2) & 37\,327.644 & 1.373 & -3.307 & 0.045 & 2.455 & 0.368 & 0.279 & 1.213 & -0.031 & 0.055 & -0.422 & -0.398 & 37\,328.46(13) \\
(1,3) & 34\,946.747 & 1.288 & -3.144 & 0.050 & 2.304 & 0.346 & 0.258 & 1.103 & -0.029 & 0.052 & -0.392 & -0.369 & 34\,947.48(12) \\
(1,4) & 32\,654.638 & 1.205 & -2.979 & 0.055 & 2.154 & 0.325 & 0.261 & 1.020 & -0.027 & 0.049 & -0.364 & -0.342 & 32\,655.32(11) \\
(1,5) & 30\,437.196 & 1.127 & -2.813 & 0.058 & 2.010 & 0.305 & 0.239 & 0.925 & -0.025 & 0.046 & -0.337 & -0.316 & 30\,437.80(11) \\
(1,6) & 28\,280.421 & 1.049 & -2.645 & 0.059 & 1.858 & 0.283 & 0.220 & 0.824 & -0.024 & 0.044 & -0.312 & -0.292 & 28\,280.95(10) \\
(1,7) & 26\,170.618 & 0.971 & -2.474 & 0.060 & 1.709 & 0.261 & 0.201 & 0.727 & -0.022 & 0.041 & -0.287 & -0.268 & 26\,171.08(9)  \\
(1,8) & 24\,093.944 & 0.895 & -2.300 & 0.060 & 1.553 & 0.240 & 0.182 & 0.629 & -0.020 & 0.038 & -0.262 & -0.245 & 24\,094.33(8)  \\
(1,9) & 22\,036.009 & 0.819 & -2.122 & 0.058 & 1.370 & 0.219 & 0.163 & 0.508 & -0.019 & 0.035 & -0.238 & -0.222 & 22\,036.29(7)  \\ [1mm]
(2,0) & 42\,162.530 & 1.542 & -3.601 & 0.027 & 2.733 & 0.412 & 0.336 & 1.447 & -0.034 & 0.060 & -0.481 & -0.456 & 42\,163.52(15) \\
(2,1) & 39\,571.598 & 1.451 & -3.440 & 0.036 & 2.579 & 0.388 & 0.311 & 1.326 & -0.032 & 0.057 & -0.448 & -0.424 & 39\,572.50(14) \\
(2,2) & 37\,099.164 & 1.364 & -3.279 & 0.043 & 2.425 & 0.365 & 0.287 & 1.207 & -0.031 & 0.054 & -0.418 & -0.394 & 37\,099.98(13) \\
(2,3) & 34\,729.525 & 1.280 & -3.116 & 0.049 & 2.276 & 0.342 & 0.265 & 1.095 & -0.029 & 0.052 & -0.388 & -0.366 & 34\,730.25(12) \\
(2,4) & 32\,447.862 & 1.199 & -2.953 & 0.053 & 2.126 & 0.316 & 0.242 & 0.984 & -0.027 & 0.049 & -0.360 & -0.339 & 32\,448.51(11) \\
(2,5) & 30\,240.020 & 1.120 & -2.788 & 0.056 & 1.981 & 0.302 & 0.239 & 0.910 & -0.025 & 0.046 & -0.334 & -0.313 & 30\,240.62(10) \\
(2,6) & 28\,092.116 & 1.041 & -2.621 & 0.058 & 1.832 & 0.281 & 0.221 & 0.813 & -0.023 & 0.043 & -0.308 & -0.289 & 28\,092.64(10) \\
(2,7) & 25\,990.449 & 0.964 & -2.451 & 0.059 & 1.682 & 0.260 & 0.203 & 0.717 & -0.022 & 0.040 & -0.283 & -0.265 & 25\,990.90(9)  \\
(2,8) & 23\,921.136 & 0.889 & -2.277 & 0.058 & 1.529 & 0.239 & 0.184 & 0.622 & -0.020 & 0.037 & -0.259 & -0.242 & 23\,921.52(8)  \\
(2,9) & 21\,869.840 & 0.813 & -2.100 & 0.057 & 1.373 & 0.217 & 0.166 & 0.527 & -0.018 & 0.034 & -0.235 & -0.219 & 21\,870.15(7)  \\ [1mm]
(3,0) & 41\,786.644 & 1.528 & -3.558 & 0.025 & 2.685 & 0.407 & 0.335 & 1.423 & -0.034 & 0.059 & -0.474 & -0.449 & 41\,787.62(15) \\
(3,1) & 39\,215.192 & 1.438 & -3.398 & 0.034 & 2.532 & 0.384 & 0.313 & 1.304 & -0.032 & 0.056 & -0.442 & -0.417 & 39\,216.08(14) \\
(3,2) & 36\,760.783 & 1.352 & -3.238 & 0.041 & 2.382 & 0.362 & 0.290 & 1.189 & -0.030 & 0.054 & -0.411 & -0.388 & 36\,761.58(13) \\
(3,3) & 34\,407.831 & 1.267 & -3.077 & 0.047 & 2.234 & 0.340 & 0.269 & 1.081 & -0.028 & 0.051 & -0.382 & -0.360 & 34\,408.55(12) \\
(3,4) & 32\,141.595 & 1.188 & -2.914 & 0.051 & 2.086 & 0.318 & 0.247 & 0.975 & -0.027 & 0.048 & -0.355 & -0.333 & 32\,142.24(11) \\
(3,5) & 29\,947.980 & 1.109 & -2.750 & 0.054 & 1.942 & 0.298 & 0.238 & 0.891 & -0.025 & 0.045 & -0.328 & -0.308 & 29\,948.56(10) \\
(3,6) & 27\,813.188 & 1.031 & -2.584 & 0.056 & 1.795 & 0.278 & 0.220 & 0.795 & -0.023 & 0.043 & -0.303 & -0.284 & 27\,813.70(9)  \\
(3,7) & 25\,723.515 & 0.955 & -2.416 & 0.056 & 1.647 & 0.257 & 0.203 & 0.702 & -0.022 & 0.040 & -0.279 & -0.260 & 25\,723.96(9)  \\
(3,8) & 23\,665.107 & 0.880 & -2.244 & 0.056 & 1.495 & 0.236 & 0.185 & 0.609 & -0.020 & 0.037 & -0.255 & -0.238 & 23\,665.48(8)  \\
(3,9) & 21\,623.545 & 0.804 & -2.067 & 0.055 & 1.340 & 0.216 & 0.167 & 0.514 & -0.018 & 0.034 & -0.231 & -0.215 & 21\,623.84(7)  \\ [1mm]
(4,0) & 41\,294.193 & 1.510 & -3.501 & 0.022 & 2.624 & 0.401 & 0.332 & 1.389 & -0.033 & 0.058 & -0.465 & -0.440 & 41\,295.14(15) \\
(4,1) & 38\,748.286 & 1.421 & -3.343 & 0.031 & 2.473 & 0.379 & 0.311 & 1.273 & -0.032 & 0.056 & -0.433 & -0.409 & 38\,749.15(14) \\
(4,2) & 36\,317.502 & 1.335 & -3.184 & 0.038 & 2.326 & 0.357 & 0.290 & 1.161 & -0.030 & 0.053 & -0.403 & -0.380 & 36\,318.28(13) \\
(4,3) & 33\,986.398 & 1.252 & -3.025 & 0.044 & 2.180 & 0.335 & 0.271 & 1.057 & -0.028 & 0.050 & -0.375 & -0.353 & 33\,987.10(12) \\
(4,4) & 31\,740.365 & 1.172 & -2.864 & 0.048 & 2.034 & 0.314 & 0.249 & 0.953 & -0.026 & 0.047 & -0.347 & -0.326 & 31\,740.99(11) \\
(4,5) & 29\,565.382 & 1.094 & -2.702 & 0.051 & 1.891 & 0.294 & 0.236 & 0.864 & -0.025 & 0.045 & -0.321 & -0.301 & 29\,565.94(10) \\
(4,6) & 27\,447.714 & 1.017 & -2.537 & 0.053 & 1.747 & 0.273 & 0.218 & 0.771 & -0.023 & 0.042 & -0.296 & -0.278 & 27\,448.21(9)  \\
(4,7) & 25\,373.700 & 0.942 & -2.371 & 0.054 & 1.599 & 0.253 & 0.201 & 0.679 & -0.021 & 0.039 & -0.272 & -0.254 & 25\,374.12(8)  \\
(4,8) & 23\,329.495 & 0.868 & -2.200 & 0.053 & 1.452 & 0.233 & 0.184 & 0.589 & -0.020 & 0.036 & -0.249 & -0.232 & 23\,329.85(8)  \\
(4,9) & 21\,300.601 & 0.793 & -2.025 & 0.052 & 1.298 & 0.212 & 0.166 & 0.496 & -0.018 & 0.033 & -0.225 & -0.210 & 21\,300.89(7)  \\
\hline
\end{tabular}
\caption{Numerical results for the spin-orbit coefficient $c_e$ in H$_2^+$ for the  range of rovibrational states $(L=1-4)$ and $(v=0-9)$ (in kHz). All definitions are identical to those given in Table~III. \label{cetable-h2plus-extended}}
\end{table}

\begin{table} [h!]
\small
\begin{tabular}{|@{\hspace{1mm}}c@{\hspace{1mm}}|@{\hspace{1mm}}c@{\hspace{1mm}}|@{\hspace{1mm}}c@{\hspace{1mm}}|@{\hspace{1mm}}c@{\hspace{1mm}}
|@{\hspace{1mm}}c@{\hspace{1mm}}|@{\hspace{1mm}}c@{\hspace{1mm}}|@{\hspace{1mm}}c@{\hspace{1mm}}|@{\hspace{1mm}}c@{\hspace{1mm}}|}
\hline
$(L,v)$ & $\displaystyle d_1^{(BP)}$ & $\displaystyle\mathcal{U}_{W}^{(2)}$&$\displaystyle\mathcal{U}_{CM}^{(2)}$ & $\displaystyle\Delta E_{ss\hbox{-}H_B}^{(2)}$ & $\displaystyle\Delta d_1^{(6)}$ &  $\displaystyle\Delta d_1^{(7\ln)} $ & $\displaystyle d_1$ (this work) \\[2mm]
\hline
(1,0) & 8565.983 & -0.802 & 0.092 & 0.951 & 0.241 & -0.050 & 8566.174(17) \\
(1,1) & 8022.434 & -0.761 & 0.089 & 0.893 & 0.222 & -0.047 & 8022.609(16) \\
(1,2) & 7505.293 & -0.721 & 0.086 & 0.837 & 0.203 & -0.044 & 7505.452(15) \\
(1,3) & 7011.264 & -0.681 & 0.082 & 0.780 & 0.182 & -0.041 & 7011.406(14) \\
(1,4) & 6537.247 & -0.642 & 0.079 & 0.740 & 0.178 & -0.039 & 6537.386(13) \\
(1,5) & 6080.286 & -0.603 & 0.076 & 0.676 & 0.149 & -0.036 & 6080.400(12) \\
(1,6) & 5637.523 & -0.564 & 0.072 & 0.629 & 0.137 & -0.033 & 5637.627(11) \\
(1,7) & 5206.141 & -0.525 & 0.068 & 0.580 & 0.122 & -0.031 & 5206.233(10) \\
(1,8) & 4783.309 & -0.486 & 0.063 & 0.523 & 0.100 & -0.028 & 4783.381(9) \\
(1,9) & 4366.125 & -0.447 & 0.059 & 0.479 & 0.091 & -0.026 & 4366.190(9) \\ [1mm]
(3,0) & 940.8385 & -0.087 & 0.010 & 0.103 & 0.0259 & -0.0054 & 940.8590(18) \\
(3,1) & 881.0351 & -0.083 & 0.010 & 0.097 & 0.0239 & -0.0051 & 881.0539(17) \\
(3,2) & 824.1126 & -0.078 & 0.010 & 0.091 & 0.0218 & -0.0048 & 824.1296(16) \\
(3,3) & 769.7077 & -0.074 & 0.009 & 0.085 & 0.0196 & -0.0045 & 769.7227(15) \\
(3,4) & 717.4796 & -0.070 & 0.009 & 0.079 & 0.0177 & -0.0042 & 717.4931(14) \\
(3,5) & 667.1019 & -0.066 & 0.008 & 0.073 & 0.0160 & -0.0039 & 667.1140(13) \\
(3,6) & 618.2585 & -0.061 & 0.008 & 0.068 & 0.0142 & -0.0036 & 618.2691(12) \\
(3,7) & 570.6370 & -0.057 & 0.008 & 0.063 & 0.0128 & -0.0033 & 570.6465(11) \\
(3,8) & 523.9225 & -0.053 & 0.007 & 0.057 & 0.0112 & -0.0031 & 523.9306(10) \\
(3,9) & 477.7905 & -0.048 & 0.006 & 0.052 & 0.0096 & -0.0028 & 477.7973(9)  \\
\hline
\end{tabular}
\caption{Numerical results for the spin-spin tensor coefficient $d_1$ in H$_2^+$ with  range of rovibrational states $(L=1-4)$ and $(v=0-9)$ (in kHz). All definitions are identical to those given in Table~IV. \label{d1results-extended}}
\end{table}

\section{Other coefficients of the effective spin Hamiltonian} \label{hfs-coeff}

\begin{table}[h!]
\begin{tabular}{|p{1cm}|p{1.9cm}|p{1.25cm}|p{1.25cm}|}
 \hline
$(L,v)$ & $b_F$       & $c_I$   & $d_2$   \\
\hline
$(1,4)$ & 836 728.705 & -35.826 & -16.414 \\
$(1,5)$ & 819.226 705 & -34.148 & -15.531 \\
$(1,6)$ & 803 174.518 & -32.385 & -14.633 \\
\hline
\end{tabular}
\caption{Hyperfine coefficients for a few rovibrational states of H$_2^+$ (in kHz). The value of $b_F$ (resp.~$c_I$, $d_2$) is taken from~\cite{Karr20} (resp.~\cite{Korobov06}). Uncertainties are discussed in the main text. To match the notations of Ref.~\cite{Korobov06}, all the $d_2$ values should be multiplied by $3(2L-1)(2L+3)=15$.}
\end{table}

\section{Derivatives of hyperfine intervals with respect to the hyperfine coefficients} \label{derivatives}

\begin{table}[h!]
\begin{tabular}{ |p{1cm}|p{1.25cm}|p{1.25cm}|p{1.25cm}|p{1.25cm}|p{1.25cm}|}
\hline
$(L,v)$ & $\gamma_{c_e,v}$ & $\gamma_{c_I,v}$ & $\gamma_{b_F,v}$ & $\gamma_{d_1,v}$ & $\gamma_{d_2,v}$ \\
\hline
(1,4) & 0.488 & -1.989 & 0.0013 & -0.266 & 0.257 \\
\hline
(1,5) & 0.489 & -1.990 & 0.0012 & -0.252 & 0.244 \\
\hline
(1,6) & 0.490 & -1.991 & 0.0011 & -0.238 & 0.230 \\
\hline
\end{tabular}
\caption{Derivatives of the interval between the $(F=1/2,J=3/2)$ and $(F=1/2,J=1/2)$ states for three rovibrational levels of H$_2^+$.}
\end{table}

\end{document}